# An Axi-Symmetric Segmented Composite SKA Dish Design: Performance and Production Analysis

M.V. Ivashina[1], R. Bakker[2], J.G. bij de Vaate[3,4], O.A. Iupikov[1], M. Arts[3], J. Dekker[2], A. van Ardenne[1,3]

[1]Onsala Space Observatory, Dept. of Earth and Space Sciences, CHALMERS University of Technology,
S-41296 Gothenburg, Sweden, marianna.ivashina@chalmers.se
[2]AIRBORNE Composites B.V., P.O. Box 24031, The Hague, The Netherlands, R.Bakker@airborne.nl
[3]ASTRON, P.O. Box 2, Dwingeloo, Netherlands, vaate@astron.nl
[4]ICRAR, Curtin Institute of Radio Astronomy, GPO Box U1987, Perth, Australia

*Abstract* — A concept of an axi-symmetric dish as antenna reflector for the next generation radio telescope – the Square Kilometre Array (SKA) – is presented. The reflector is based on the use of novel thermoplastic composite material (reinforced with carbon fibre) in the context of the telescope design with wide band single pixel feeds. The baseline of this design represents an array of 100's to 1000's reflector antennas of 15-m diameter and covers frequencies from <1 to 10 GHz. The purpose of our study is the analysis of the production cost of the dish and its performance in combination with a realistic wideband feed (such as the 'Eleven Antenna' feed) over a wide frequency band and a range of elevation angles. The presented initial simulation results inidicate the potential of the proposed dish concept for low-cost and mass production and demonstrate sensitivity comparable to that of the presently considered off-set Gregorian reflector antenna with the same projected aperture area. We expect this observation to be independent of the choice of the feed, as several other single-pixel wideband feeds (that have been reported in the literature) have similar beamwidth and phase center location, both being rather constant with frequency.

## I. INTRODUCTION

The international radio astronomy community is currently pursuing the development of the next generation radio telescope known as the Square Kilometre Array (SKA), which will operate from 70 MHz to 10 GHz with unprecedentedly high sensitivity, resolution and survey speed. The SKA reference [1] design includes several novel antenna array technologies: for the lower and middle frequency bands sparse and dense aperture arrays will be used, and for high frequency range, i.e. from 1 to 10 GHz, reflector antennas fed with multi-beam Phased Array Feeds (PAF) and/or wide band Single Pixel Feeds (SPFs) such as described in [2-3] will be used. At present, the SKA will be constructed in two phases: phase I ($SKA_1$) and phase II ($SKA_2$). $SKA_1$ will have unique science capabilities with much lower investment levels and less collecting areas as the final SKA. The $SKA_1$ dish array will have the sensitivity of up to 1000 $m^2$/K using approximately two hundred and fifty 15-metre antennas, employing a SPF instrumentation package. $SKA_2$ will have a sensitivity of 10,000 $m^2$/K realized with about 2500 15-metre antennas.

This paper focuses on the analysis of the production of the dish and its performance in combination with a realistic wideband SPF feed, such as for example the so-called Eleven antenna [3]. Although in principle feasible, the present study does not investigate extensions and/or alternative designs incorporating PAF systems and multi-combinations of feeds such as described in [4-5]. An axis-symmetric dish antenna reflector concept has been selected. In general, the symmetric dish concept is attractive because it's design is relatively straightforward while wide spread knowledge and performance experience is available. The novelty involves the reflector structure material, which is based on carbon re-enforced thermoplastic material, not (yet) used for this kind of applications and looking highly promising and feasible. It is of interest to note that other markets e.g. automotive and aerospace are interested in applying this material, pushing further developments.

## II. EM-ANALYSIS OF SOLID REFLECTOR DESIGNS WITH THE ELEVEN FEED

### A. The Axi-Symmetric Reflector Antennas

The starting point for this analysis is the EM-modeling of the 15-metre antenna in combination with the Eleven feed to determine the optimal geometry that will lead to the maximum system sensitivity. This modeling procedure has been carried out in the TICRA's software GRASP9.4 by using the measured far-field patterns of the Eleven antenna. For this purpose, the measured field data [3] was converted to the GRASP spherical cut format which describes the field with a number of points defined in 360 $\varphi$-cuts on a sphere with 181 points in each cut. The simulations were performed for several frequencies within the operational range of the Eleven antenna (2 - 12 GHz). Afterwards, the simulated patterns of the total feed-reflector antenna system were used to compute the aperture efficiency of the antenna $\eta_{ap}$ and the noise temperature contribution $T_a$ due to the ground noise caused by spillover effects. The noise temperature was evaluated for different antenna pointing

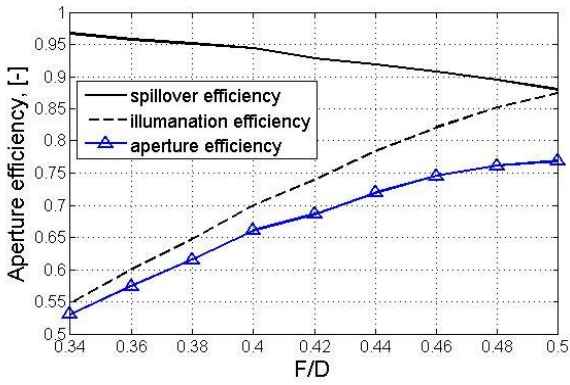

(a) The aperture efficiency of the antenna versus the focal length to diameter ratio of the reflector F/D.

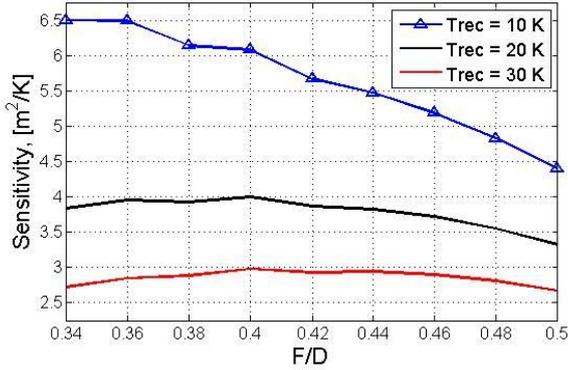

(b) The sensitivity ($A_{ph}/T_{sys}$), where $T_{sys}$ is defined as a sum of the receiver noise temperature (assumed to be equal to 10, 20 and 30 K) and the antenna noise temperature contribution due to the ground noise pick up that was averaged over the range of antenna pointing directions for $El$=10-170°.

Fig. 1 The simulated performance parameters of the axi-symmetric reflector antenna with F/D=0.4 (half subtended angle of 60°) when fed with the Eleven feed at 5.6 GHz, based on the measured feed patterns.

elevation angles, while the azimuth angle was set to a constant value of 0°. Elevation angle ($El$) is defined as the pointing angle up from the ground or horizon, $El$=90° is the 'zenith' pointing angle. Often in the reflector antenna and feed designs, the elevation angle is assumed to be constant and equal to 90° However for low noise radio astronomy applications this assumption neglects the pointing direction dependent ground noise pick up [6]. It is therefore essential to include $El$ as a parameter in the simulation scenarios of radio telescopes.

Fig.1 presents the first set of simulation results for the axi-symmetric reflector with F/D=0.34-0.5 for the case when the antenna operates at 5.6 GHz. Fig.1 (a) shows the aperture efficiency including its spillover and illumination components and Fig.1 (b) illustrates the sensitivity ($\eta_{ap} A_{ph}/T_{sys}$), where $T_{sys}=T_a+T_{rec}$, based on the assumed receiver noise temperature $T_{rec}$ of 10K, 20K and 30K and the antenna noise temperature $T_a$ averaged over several antenna pointing directions within the range of $El$=10-170°. These results demonstrate there are optimal F/D ratios of the prime-focus reflector antenna that maximize the system sensitivity.

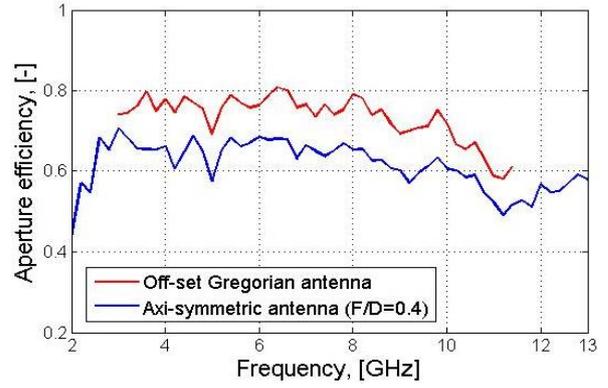

(a) The aperture efficiencies of the reflector antennas versus frequency

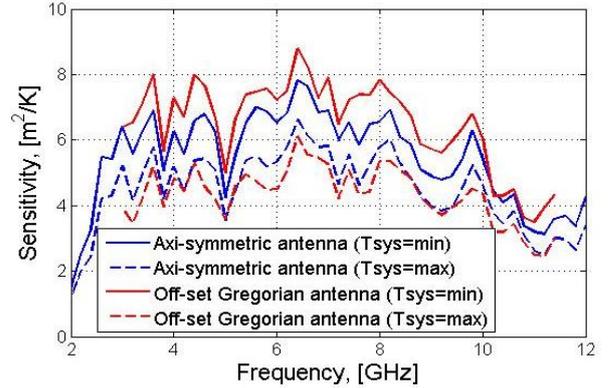

(b) The sensitivity ($A_{ph}/T_{sys}$), with the assumed constant over frequency receiver noise temperature of 10K. The upper and lower bounds of sensitivity correspond to the minimum and maximum values of the antenna ground noise contribution over the range of elevation angles ($El$=10-170°).

Fig. 2 The Comparison of the simulated performance parameters for the two considered reflector antennas, based on the measured Eleven feed patterns.

This maximum is the result of trade-off between high antenna efficiency of shallow- and low noise performance of deep reflector antennas. For the cooled and un-cooled receiver systems with $T_{rec}$ ranging from 20 to 30K [7-8], the optimal F/D is equal to 0.40. For very high performance receivers with $T_{rec}$<10K, the optimal F/D value is smaller (0.34) due to comparable contributions of $T_a$ and $T_{rec}$ to the total system noise. Since, the beam shape of the Eleven antenna is virtually frequency invariant (except for very low and very high frequencies of the operational band), the optimal reflector subtended angle is only weakly dependent of frequency. In a further analysis, we will now consider the baseline design of the antenna with F/D=0.4 that corresponds to the reflector half subtended angle of 60°.

### B. Comparison with the SKA Off-Set Gregorian Reflector Antenna Design:

Fig.2 (a) and (b) compare, respectively, the aperture efficiency and sensitivity of the selected axi-symmetric antenna and the SKA off-set Gregorian reflector design (with the sub-reflector half subtended angle of 55°)[9], also equipped with the Eleven feed and modelled using the same simulation tool and analysis approach as described in Sec.II.A. As one can see on fig.2 (a) the efficiency of the axi-symmetric antenna is equal to 50-70% in the wide frequency band (2-13 GHz) and, as expected, approximately 10-15% lower as compared to the dual-reflector system. The antenna noise temperatures (not shown here) have been found to vary from 7 to 20K and from 10 to 30K, respectively for the symmetric and off-set designs, depending on frequency and elevation angle $El$. The average value of $T_a$ over elevation and frequency for the prime-focus antenna is 10K in the frequency band 3.5-10.5GHz. For the Gregorian design it is higher (15K) due to the stronger elevation dependence of $T_a$, and in particularly its increase for off-zenith pointing directions (except for $El$=40-80°). Fig.2(b) compares the sensitivities of two antenna systems over the frequency band for $T_{rec}$=10K, as an example. The simulations have been performed also for $T_{rec}$ of 20K and 30K and shown that the sensitivity upper bound (defined by the minimum $T_a$) is always higher for the Gregorian system, while its lower bound (corresponding to the maximum $T_a$) is mostly higher for the axi-symmetric one. The average sensitivity over elevation and frequency for the latter system is equal to 5.6m$^2$/K and comparable to 5.5 m$^2$/K of the dual-reflector design.

III. REFLECTOR DESIGN CONCEPT BASED ON THERMOPLASTIC COMPOSITE MATERIALS

A. *Material selection and structural design*

Composite materials are increasingly used for all sorts of components in different markets. The material is lightweight and has high specific performance properties, which often outperform metal equivalents, resulting in a lightweight design. With a composite solution, weight can typically be reduced with 30% compared to a steel / aluminium option. Additionally, this is beneficial for transport of the reflector, material costs, pedestal design, and required actuators that the weight of the reflector itself is kept to a minimum.

Furthermore thermoplastic composite material can be shaped with fibre placement, thermoforming; welding and conventional cutting operations and all processes can be automated. With this concept the aim is to make a "step change" in production efficiency, as the thermoplastic material (light weight reflector construction) will be used in combination with the new automated, industrialised production technologies.

The baseline design of the reflector uses a stiffened skin with several different stiffeners. The entire structure will be build from the same raw material, which is a thermoplastic carbon based composite.

All stiffeners will be built up from relative simple parts which are welded together. T-stiffeners are more efficient (higher bending stiffness per mass) compared to the blade-stiffeners, it is however also more difficult to produce. Therefore, the T-stiffener is selected for the more critical

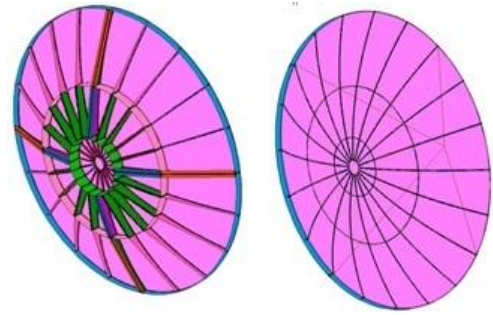

Fig. 3 Symmetric reflector design

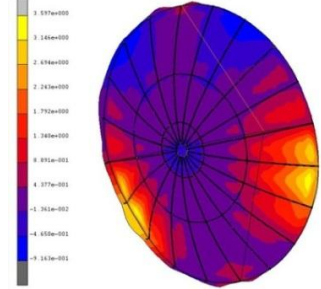

Fig. 4 Displacement results with 4m/s wind at 120° elevation

and higher loaded regions, whereas the blade-stiffener is used in regions with lower requirements.

Within the frame structure also 4 tangential stiffeners are added to increase reflector performance. These rings can be seen in fig.3. The whole self supporting 15 meter reflector structure as shown has a weight of 1580kg and the design still leaves room for further optimisation in weight, performance and cost.

The current design is analysed using a Finite Element Analysis (FEA) with MSC Marc & Mentat 2010. The Coefficient of Thermal Expansion (CTE) properties of the composite material are mostly dependent on the CTE properties of the carbon fibres, which are low compared to other frequently used antenna materials. This results in a thermal stable construction, see the results in Table 1.

Several thermal load cases have been defined, with temperatures as high as 70 °C and a temperature difference between skin and frame of up to 20 degrees. Wind loads are defined up to 12 m/s [10-11] and gravity is taken into account on the whole reflector movement range.

Table I represents single or combinations of average load cases. When the extreme load cases that have been specified in the requirements are combined, which is not a likely to occur simultaneously, the performance of the reflector design still has a rms of 0,93mm.

All parts of the reflector structure are manufactured separately and then assembled into one large structure by welding the thermoplastic material together. Production of the – many identical – parts can be done by fully automated production processes using among others press forming. The design with assembly of subsection also gives the opportunity for optimising transport and final installation, as it would even be possible to assemble the reflector in a few sub sections and weld them together on-site.

TABLE I
RMS VALUES FOR TYPICAL LOAD CASES

| Description | RMS fitted surface |
|---|---|
| Gravity 15° elevation | 0.25 |
| Gravity 60° elevation | 0.34 |
| Gravity 90° elevation | 0.44 |
| Typical thermal load case (33°C skin and 37°C rib structure) | 0.04 |
| Typical wind (4 [m/s]), 60° elevation | 0.06 |
| Typical wind (4 [m/s]),120° elevation | 0.02 |
| Total typical (thermal + wind + gravity, 60° elevation) | 0.41 |
| Manufacturing and assembly accuracy | 0.5 |
| Total typical incl. manufacturing and assembly | 0.65 |

*B. Reflectivity and segment impact analysis*

For optimum performance a metal mesh is embedded in the surface of the skin material. Several combinations of composites and metal meshes where tested to optimise for lowest noise temperature. We investigated the effect of gaps between the segments on the antenna noise temperature $T_{leak}$ due to the ground noise pick up through the mesh (ground radiation leakage). For this study, we have simulated (using GRASP9) the antenna patterns with the solid and segmented dishes and then computed $T_{leak}$ (assuming the sky noise temperature $T_{sky}$=0 K and $T_{ground}$=300 K). The blue dot in fig. 5 indicates the result ($T_{leak}$=0.1 K) for the solid dish. The blue line shows the ground noise leakage for a panelled dish as a function of the gap area to dish area ratio $A_{gaps}/A_{solid}$. One can see that even for small gaps the ground noise contribution is significantly higher (around 0.7 K). The red line shows the function $T_{leak}$=300($A_{gaps}/A_{solid}$) which gives an approximation of the simulated in GRASP result that seems to be reasonable for $A_{gaps}/A_{solid}$ larger than ~0.01.

IV. CONCLUSIONS

While recognizing that detailed study needs to be done to advance this work, the presented results indicate that the axi-symmetric thermocomposite reflector represents an attractive solution in terms of its potential for low cost and suitability for mass production for the SKA design, while its sensitivity performance (when fed with the Eleven feed) is comparable to that of the presently considered (and likely more expensive) off-set Gregorian antenna designs. In future, it would be useful to perform this comparison with other wideband feeds such as a quad-ridged flared horn and non-planar quasi-self-complementary feed. Other important questions include polarimetry and suitability for deep imaging and wide-field radio astronomy observations.

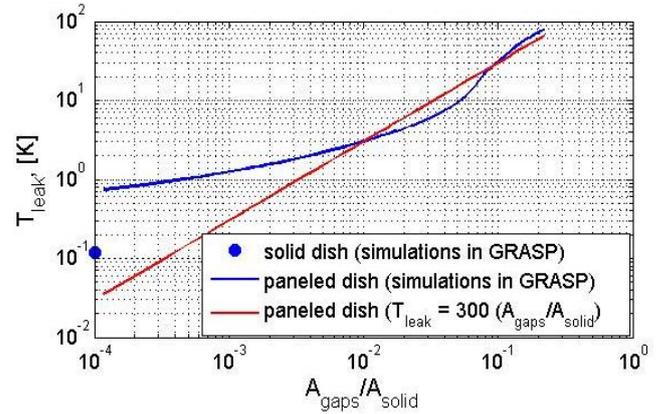

Fig. 5. The antenna noise temperature $T_{leak}$ [K] due to the ground noise pick up through the mesh versus gap area between segments.

ACKNOWLEDGEMENTS

The authors wish to acknowledge Dr. J.Yang for providing us with the Eleven feed measured patterns and Dr. L. Baker (Cornell) for the details of the SKA Gregorian reflector design. The study in Sec.II was partly supported by the Swedish Agency for Innovation Systems VINNOVA.